# Anomalous Characteristics of the Generation –Recombination Noise in Quasi-One-Dimensional Van der Waals Nanoribbons

**Adane K. Geremew[1], Sergey Rumyantsev[1,2], Matthew A. Bloodgood[3], Tina T. Salguero[3] and Alexander A. Balandin[1,*]**

[1]Nano-Device Laboratory (NDL) and Phonon Optimized Engineered Materials (POEM) Center, Department of Electrical and Computer Engineering, University of California, Riverside, California 92521 USA

[2]Ioffe Physical-Technical Institute, St. Petersburg 194021 Russia

[3]Department of Chemistry, University of Georgia, Athens, Georgia 30602 USA

## Abstract

We describe the low-frequency current fluctuations, *i.e.* electronic noise, in quasi-one-dimensional ZrTe$_3$ van der Waals nanoribbons, which have recently attracted attention owing to their *extraordinary* high current carrying capacity. Whereas the low-frequency noise spectral density, $S_I/I^2$, reveals $1/f$ behavior near room temperature, it is dominated by the Lorentzian bulges of the generation–recombination noise at low temperatures ($I$ is the current and $f$ is the frequency). Unexpectedly, the corner frequency of the observed Lorentzian peaks shows strong sensitivity to the applied source–drain bias. This dependence on electric field can be explained by the Frenkel–Poole effect in the scenario where the voltage drop happens predominantly on the defects, which block the quasi-1D conduction channels. We also have found that the activation energy of the characteristic frequencies of the G-R noise in quasi-1D ZrTe$_3$ is defined primarily by the temperature dependence of the capture cross-section of the defects rather than by their energy position. These results are important for the application of quasi-1D van der Waals materials in ultimately downscaled electronics.

* Corresponding author (A.A.B.): balandin@ece.ucr.edu ; web-site: http://balandingroup.ucr.edu/

Two-dimensional (2D) materials, such as graphene and transition metal dichalcogenides (TMDs) $MX_2$ (where M = many transition metals; X = S, Se, Te), have revealed new physics and demonstrated potential for practical applications [1–11]. In recent years, interest in layered van der Waals materials has expanded to include quasi-one-dimensional (1D) structures and compositions. Unlike the layered $MX_2$ materials that yield 2D nanometer thickness sheets upon exfoliation, the transition metal trichalcogenides (TMTs) $MX_3$ contain 1D motifs, *i.e. atomic threads*, that are weakly bound together by van der Waals forces. Examples include $TiS_3$, $NbS_3$, and $TaSe_3$ [12–14]. As a consequence of their structures, the mechanical exfoliation of $MX_3$ crystals results in nanowires and nanoribbons rather than 2D layers. We have discovered that some of these quasi-1D nanomaterials reveal an exceptionally high current density. For instance, quasi-1D $TaSe_3$ nanowires capped with *h*-BN layers have a breakdown current density exceeding $J_B \sim 10$ MA/cm$^2$, which is larger than what can be sustained by the state-of-the-art Cu interconnects [15]. In a recent contribution, we reported uncapped $ZrTe_3$ nanoribbons with an even more impressive breakdown current density of $J_B \sim 100$ MA/cm$^2$ [16], which is more than an order of magnitude larger than the value obtained in DC testing of Cu wires [17,18].

In this Letter, we report results pertaining to the low-frequency current fluctuations, *i.e.* electronic noise, in $ZrTe_3$ nanoribbons. The low-frequency noise is a ubiquitous phenomenon, present in all kinds of electronic materials and devices [19–21]. Practical applications, even for high-frequency devices, require the reduction of low-frequency noise to an acceptable level, due to possible up-conversion of the low-frequency noise to the phase and amplitude noise of the high-frequency devices. Additionally, the specific features of the low-frequency current fluctuations can provide valuable information about electronic transport, typical defects, grain boundaries, and charge carrier recombination dynamics [20,22–26]. The frequency, bias and temperature dependences of the low-frequency noise spectral density have been used as reliability metrics for devices and interconnects by the electronics industry [27–30]. For these reasons, our study is important for the proposed applications of quasi-1D $ZrTe_3$ nanoribbons in ultimately downscaled device channels and interconnects. From the fundamental science point of view, characterization of the low-frequency current fluctuations can shed light on the electron transport properties of $ZrTe_3$. We have focused our studies to room temperature (RT) and below to elucidate the physical mechanism of the current fluctuations and determine possible activation energies for various processes contributing to the noise. The results indicate that the normalized low-frequency noise spectral density, $S_I/I^2$, reveals $1/f$ behavior near RT ($I$ is the current and $f$ is the frequency). However, at lower temperatures, the noise spectral density is dominated by the Lorentzian bulges of the generation–recombination (G-R) noise. Interestingly, the corner frequency of the observed Lorentzian peaks shows strong sensitivity to the applied source–drain bias.

Similar to the other group IV $MX_3$ materials, $ZrTe_3$ crystallizes in the monoclinic space group $P2_1/m$ with $Z = 2$. Early work suggested the existence of $ZrTe_3$ polymorphs ("A" and "B"). However, later studies confirmed only the original crystal structure ("A") [31–34]. As illustrated in Figure 1a, this structure is composed of pairs of trigonal prismatic $ZrTe_3$ columns extending along the $b$-axis. These columns are stacked in layers along the $c$-axis, which are weakly bound by van der Waals forces. An additional unique feature is significant Te–Te⋯Te–Te interactions along the $a$-axis. As a result, $ZrTe_3$ exhibits more 2D character than many other $MX_3$ compositions and thus can be more completely considered a quasi-1D/2D material.

[Figure 1]

The intriguing properties of $ZrTe_3$ at low temperatures have been studied extensively. The bulk material is metallic with a charge density wave (CDW) transition at 63 K and a superconducting (SC) transition at 2 K [35]. The electrical responses caused by the CDW and filamentary SC states occur primarily along the $a$-axis, $i.e.$ perpendicular to the $ZrTe_3$ columns, due to the pronounced Te–Te⋯Te–Te interactions. Numerous reports describe the effects of pressure [36,37], magnetic field [38], metal intercalation (e.g., $Ni_{0.01}ZrTe_3$) [39], and chemical substitution (e.g., $ZrTe_{3-x}Se_x$) [40] on the CDW and SC properties of $ZrTe_3$. Notably, these studies were all based on bulk samples, thus the impact of nanostructuring on the properties of $ZrTe_3$ is not yet understood. Nanostructure – substrate interactions, strain, defects, electron and phonon confinement – all can affect resistivity dependence on temperature, and the type of electrical conduction, $e.g.$ metallic $vs.$ semiconducting.

For this study, crystals of $ZrTe_3$ were synthesized by iodine-mediated chemical vapor transport (CVT). Details regarding material synthesis are provided in the Supplemental. The crystal structure was confirmed by single crystal X-ray diffraction; the resulting high-quality dataset provided a unit cell and atomic coordinates (Tables S1 and S2) that are in excellent agreement with type "A" $ZrTe_3$. Nanoribbons were obtained by the standard mechanical exfoliation method. The quality of these nanoribbons was assessed by high-resolution transmission electron microscopy (HRTEM). As shown in Figure 1b-c, HRTEM reveals large regions of defect-free $ZrTe_3$; the atomic distances are consistent with the array of $ZrTe_3$ columns illustrated in Figure 1a. Selected area electron diffraction (SAED) (Figure 1d) also is consistent with single crystalline nanoribbon quality. Furthermore, the material composition was evaluated by energy dispersive spectroscopy (EDS) (Figure S1) and Raman spectroscopy[16].

The devices were fabricated by the shadow mask method on the mechanically exfoliated quasi-1D nanoribbons placed on a $Si/SiO_2$ wafer [16]. Owing to a direct deposition of the metal contacts onto the pre-selected $ZrTe_3$ nanoribbons, the shadow mask method allows one to avoid the damage and chemical contamination associated with the conventional lithographic lift-off processes. It also drastically reduces the total air exposure time during the fabrication. The shadow masks were fabricated using the double-side polished Si wafers with 3 μm thermally grown $SiO_2$ (Ultrasil Corp.; 500-μm thickness; p-type; <100>). The details of the shadow mask fabrication are provided in the Methods section. The contact metals Ti/Au (10:100 nm) were evaporated through a shadow mask, forming a testing device structure. The devices were transferred to another vacuum system for the electronic transport measurements. Figures 2 (a) and (b) show the optical microscopy of exfoliated nanoribbons of $ZrTe_3$ on $Si/SiO_2$ substrate and the atomic force microscopy (AFM) image of the fabricated device structure (device dimensions are approximately 50 nm × 200 nm), respectively. In Figure 2 (c), we present the $ZrTe_3$ channel resistance, $R$, as a function of temperature, $T$. The inset shows the low-field current-voltage (I-V) characteristics of the device. The low-field I-Vs confirm the Ohmic characteristics of the contacts. The multi-gate design of the device under test allowed us to extract the contact resistance, $R_C$, confirming that it is negligible compared to the channel resistance: $2R_C \leq 1.1\% \times R_T$, where the total resistance $R_T=R+2R_C$. As the temperature increases from $T$=100 K to $T$=300 K, the resistance of ZrTe$_3$ nanoribbon decreases, suggesting semiconducting behavior. Although bulk ZrTe$_3$ has been described previously as metallic or semi-metallic [33,35,41], [42], a band gap may open due to tensile strain or other electronic effects related to nanostructuring. Notably, NbSe$_3$ also exhibits a dramatic change from metallic behavior in the bulk state to nonmetallic in nanowire form [43,44].

[Figure 2]

The low-frequency noise measurements were performed in the temperature range from 77 K to 298 K under high vacuum. The devices were biased with a silent battery and a potentiometer biasing circuit and measured using a low-noise amplifier and a spectrum analyzer. Details of our experimental procedures, in the context of different devices, were reported by us elsewhere [45–50]. The noise spectra were acquired at a low bias voltage to avoid Joule heating. Figure 3 (a) shows the normalized noise spectral density, $S_I/I^2$, as a function of frequency at different temperatures. Since the contact resistance of the tested device was negligible, the noise response is dominated by the channel. The main observation is that at low temperatures the noise spectrum consists of the bulges of the generation – recombination (G-R) noise, typical for semiconductors. As the temperature increases, the bulges shift to a higher frequency, eventually disappearing or moving outside of the examined frequency range, when temperature approaches RT. Near RT, the noise spectrum becomes $1/f^\gamma$ type with $\gamma\approx1$, which is characteristic for both metals and semiconductors as well as majority of electronic devices [25]. In Figure 3 (b), we present $S_I/I^2\times f$ as a function of temperature in order to compensate for $1/f$ noise background and make the G-R peak shift with the temperature more visible. The overall noise level is rather small. The normalized noise spectral density, $S_I/I^2\approx10^{-10}$ Hz$^{-1}$ at $f$=10 Hz and temperature T=280 K. The existence of the G-R noise is a separate prove that studied quasi-1D ZrTe$_3$ nanoribbons demonstrate semiconducting behavior within the given temperature range.

[Figure 3 (a-b)]

In semiconductors, G-R noise is observed at low frequency and its spectral density is described by the Lorentzian: $S_I(f) = S_0/[1 + (2\pi f\tau)^2]$, where $S_0$ is the frequency independent portion of $S_I(f)$ observed at $f << (2\pi\tau)^{-1}$ and $\tau$ is the time constant associated with the return to the equilibrium of the occupancy of the level. In a typical situation of the semiconductor doped with the shallow, fully ionized donor or acceptor, and another noisy deep level, the spectral density of the G-R noise is given by [51]:

$$\frac{S_I}{I^2} = \frac{4N_t}{Vn^2}\frac{\tau F(1-F)}{1+(\omega\tau)^2},$$ (1)

where $\omega = 2\pi f$, $V$ is the sample volume, $n$ is the equilibrium electron concentration for the $n$-type semiconductor, and $F$ is the trapping state occupancy function. The G-R noise time constant, $\tau$, is expressed in terms of the trapping state capture, $\tau_c$, and emission, $\tau_e$, time constants:

$$\frac{1}{\tau} = \frac{1}{\tau_c} + \frac{1}{\tau_e},$$ (2)

which are given by:

$$\tau_c = \frac{1}{\sigma v_T n},$$ (3)

and

$$\tau_e = \frac{1}{\sigma v_T N_c \exp(-\frac{E_0}{kT})}.$$ (4)

Here $\sigma$ is the capture cross section of the trap, $v_T$ is the electron thermal velocity, $N_c$ is the effective electron density of states (DOS) in the conduction band, and $E_0$ is the trap level position, relative to the conduction band. Note that Eqs. (2) - (4) are written for the n-type semiconductor. The equations for the p-type semiconductor are analogous. As seen from Eq. (4), if the emission time dominates, the characteristic time $\tau$ depends on temperature exponentially and the energy $E_0$ can be extracted from the experimental data. Usually $ln(\tau)$ or $(f_c=(1/2\pi\tau))$ is plotted as a function of $1/T$. If the temperature dependences of the thermal velocity and DOS are significant, $ln(T^2\tau)$ can be plotted as a function of $1/T$. The slope of this Arrhenius plot defines the energy $E_0$. Figure 4 shows the Arrhenius plot of the characteristics frequency $f_c$ for the experimental data shown in Figure 3 (b).

[Figure 4]

The trap activation energy, extracted from Figure 4, is $E_0 \cong 0.18$ eV. The activation energy obtained by this method is often associated with the energy level position of a given trapping state. However, this is not always the case. The caption cross-section of the trap levels, $\sigma$, often depends exponentially on temperature [26,52].

$$\sigma = \sigma_0 \exp(-\frac{E_1}{kT}). \tag{5}$$

In such cases, the procedure, described above, yields the sum of the activation energies, $E_0 + E_1$, rather than $E_0$. The activation energies $E_0$ and $E_1$ cannot be found separately in this approach. The method to find the energies $E_0$ and $E_1$ separately was proposed in Ref. [53]. This method requires to plot the spectral noise density versus temperature $T$ at a series of frequencies. The temperature dependence of the noise for each frequency has a maximum at $T=T_{max}$. As the next step, the dependence of the noise at the point of maximum, $ln(S_{max})$, is plotted versus $ln\omega$, where $\omega=2\pi f$. The slope of this dependence is defined by the energies $E_0$ and $E_1$, which can be found separately. This method was developed for the case when the semiconductor is doped with a shallow donor center, which is fully ionized at all temperatures of the experiment. The concentration of the traps at this level is high enough so that the electron concentration does not depend on temperature, and this concentration is much higher than the concertation of the noisy deep levels. As one can see from Figure 2 (c), the resistance of the sample decreases with increasing temperature, which is an indication of the free carrier concentration temperature dependence. However, while the characteristic frequency of the G-R noise changes with temperature more than two orders of magnitude the change of the resistance, and concertation, is only a factor of ×2.5. Therefore, we can neglect the temperature dependence of the resistance and concentration in our analysis and use the method of Ref. [46].

Figure 5 shows the normalized noise spectral density, $S_I/I^2$, as a function of temperature for different frequencies. As temperature increases, the noise peak shifts to higher temperatures. The blue line and data points in Figure 6 show the dependence of $ln(S_{max})$ versus $ln\omega$. The slope of this dependence $A=\delta(ln\ (S_{max}))/\delta(ln\ \omega)\cong 1.24$. In accordance with the model of Ref. [53], $A>1$ is an indication that the level is located above the Fermi level and $A=(2E_0+E_1)/E_0+E_1)$. The red line and data points in Figure 6 shows the Arrhenius plot of the invers temperature $(1/T)$ as a function of $ln\ \omega$. The frequency of each red data point in Figure 6 corresponds to $f_c$ at temperature $T_{max}$. The slope of the Arrhenius plot, $1/kT_{max}$ versus $ln\ \omega$ is $B=\delta(1/T_{max})/\delta(ln\ \omega)\cong 1/(E_0+E_1)$. The extracted activation energy from the linear fitting gives 0.19 eV, which, within the experimental error, matches the activation energy extracted from Figure 4. With the known $A$ and $B$, the activation energy of the cross-section temperature dependence and level position can be extracted as $E_1=0.144$ eV and $E_0=0.0456$ eV. These data indicate that the activation energy of the characteristic frequencies of the G-R noise peaks in quasi-1D ZrTe$_3$ nanoribbons is dominated by the activation energy of the capture cross-section temperature dependence.

[Figure 5]

[Figure 6]

We measured the noise at low temperature ($T$=77 K) as a function of the source – drain bias. Figure 7 (a) shows the normalized noise spectral density, $S_I/I^2$, as a function of frequency at different source-drain voltages, $V_{SD}$. In Figure 7 (b), we present $S_I/I^2 \times f$ in order to compensate for $1/f$ noise background and make the peak shift with the electric bias more visible. As seen from Figure 7, the characteristic frequency $f_c$ changes about three orders of magnitude when the bias voltage increases from 45 mV to 250 mV. In general, the shift of the Lorentzian peak in the noise spectrum with the applied electric field can be attributed to reduction of the impurity barrier potential in the high electric field, which is known as the Poole-Frenkel effect [54]:

$$\Delta E_{fp} = \left(\frac{q^3 F}{\pi \varepsilon_0 \varepsilon}\right). \tag{6}$$

Here $\Delta E_{fp}$ is the reduction of the barrier, $q$ is the charge of an electron, $F$ is the electric field, $\varepsilon_0$ is the permittivity of free space, and $\varepsilon$ is the relative dielectric constant. Assuming that the characteristic frequency depends on the energy exponentially, we estimated that the electric field required to shift this frequency by the three orders of magnitude is on the order of 50 kV/cm. For the 2-μm length nanoribbon in our devices, the average field in the sample does not exceed ~1.25 kV/cm. However, the specific of the quasi-1D nanoribbon structure is that a defect can increase significantly the resistivity of the individual quasi-1D chains or even completely block it. Since the resistivity in the directions perpendicular to the atomic chain is much higher, the local field at a defect can be much higher than the average one. Most of the potential drop can happen on over the spatial extend of the defect. Assuming that the defect is a 1D line we estimate the potential as $\varphi(x) \sim en(x)\ln(L_0/a)$, where $n$ is the uncompensated charge in the wire, $a$ is the diameter of the wire, and $L_0$ is the characteristic screening length which depends on the chain dimensions. The estimate is based on the gradual channel approximation [55]. The electric field at the end of the defect is given by $F \sim en/(x - d/2)$. Then the electric field in the non-conducting gap, *i.e.* spatial extend of the defect, can be roughly estimated as:

$$F \cong \left( \frac{\varphi_1}{x+d/2} - \frac{\varphi_2}{-x+d/2} \right) \frac{1}{\ln L_0/a},$$  (6)

where $\varphi_1$ and $\varphi_2$ are the potentials on the wire (which represent a nanoribbon), and $d$ is dimension of the non-conducting gap (see inset in Figure 7 (a)). From Eq. (6), the electric field in the middle of the gap, associated with the defect, is on the order of $2K(\varphi_1-\varphi_2)/d$, where $K \geq 1$ depends logarithmically on the specific geometry close to the defect. With the potential difference of 250 mV, the electric field of 50 kV/cm can be easily obtained for the gap of 100 nm. The actual defect size may be much smaller, and the electrical field, therefore, higher. This can explain the reduction of the barrier on the defect, and the resulting strong shift of the characteristic frequency as a function of bias, in accordance with the Poole-Frenkel effect. It is natural to assume that the observed strong bias dependence of the G-R noise can be a common feature of the quasi-1D crystals. It is interesting to note that the current-voltage characteristics are perfectly linear at the considered bias but the noise spectra clearly reveal this unusual effect.

In conclusion, we investigated the low-frequency electronic noise in quasi-1D ZrTe$_3$ van der Waals nanoribbons. Such nanostructures have recently attracted attention owing to their extraordinary high current carrying capacity. Whereas the low-frequency noise spectral density, reveals 1/$f$ behavior near RT, it is dominated by the Lorentzian bulges of G-R noise at low temperatures. The corner frequency of the Lorentzian peaks shows strong sensitivity to the applied source–drain bias. This dependence on electric field can be explained by the Frenkel-Poole effect if one assumes that the voltage drop mostly happens on the defects, which block the quasi-1D conduction channels. The observed strong bias dependence of the G-R noise can be a common feature of the quasi-1D crystals. We also found that the activation energy of the characteristic frequencies of the G-R noise in quasi-1D ZrTe$_3$ is primarily defined by the temperature dependence of the capture cross-section of the defects rather than by their energy position. The activation energy of the cross-section temperature dependence and level position were found to be $E_1$=144 meV and $E_0$=45.6 meV, respectively. These results are important for the proposed applications of quasi-1D van der Waals materials in ultimately downscaled electronics.

**METHODS**

**Mask and device fabrication:** ZrTe$_3$ nanoribbons were mechanically exfoliated from the bulk crystals and transferred to Si/SiO$_2$ substrate. We utilized the shadow mask method to fabricate the prototype interconnects. By allowing direct deposition of metallic contacts onto pre-selected ZrTe$_3$ nanoribbons, this method avoids the damage and chemical contamination typically associated with conventional lithographic lift-off processes, and it also drastically reduces the total air exposure time (<2 hrs., compared to 2 – 3 days for conventional lithography processes). The shadow masks were fabricated using double-side polished Si wafers with 3 µm thermally grown SiO$_2$ (Ultrasil Corp.; 500-µm thickness; P-type; <100>). The shadow mask fabrication process began with evaporation of 200 nm Chromium (Cr) on the front side of the wafer, followed by stencil mask patterning of this layer using a combination of electron beam lithography and Cr etchant (1020A). This was followed by fluorine-based reactive ion etching (RIE) to transfer the pattern to the underlying SiO$_2$. Finally, the pattern was transferred into the underlying Si substrate using deep reactive ion etching (DRIE) (Silicon Trench Etch System; Oxford Cobra Plasma Lab Model 100). The DRIE etch step was timed to break through to a large backside window that was previously defined using lithographic patterning, RIE, and DRIE. The completed shadow masks were used to fabricate ZrTe$_3$ devices by aligning them with pre-selected nanoribbons on the device substrate, clamping the aligned mask and device substrate together, and placing the clamped assembly in an electron beam evaporator (EBE) for contact deposition (10 nm Ti and 100 nm Au). The completed devices were then transferred to another vacuum chamber for electrical characterization.

**Acknowledgements:** Device fabrication and testing were supported, in part, by the Semiconductor Research Corporation (SRC) contract 2018-NM-2796: One-Dimensional Single-Crystal van-der-Waals Metals: Ultimately-Downscaled Interconnects with Exceptional Current-Carrying Capacity and Reliability. Materials synthesis and characterization were supported, in part, by the National Science Foundation (NSF) through the Emerging Frontiers of Research Initiative (EFRI) 2-DARE project: Novel Switching Phenomena in Atomic $MX_2$ Heterostructures for Multifunctional Applications (NSF EFRI-1433395). A.A.B. also acknowledges the UC - National Laboratory Collaborative Research and Training Program - University of California Research Initiatives LFR-17-477237. Nanofabrication was performed in the Center for Nanoscale Science and Engineering (CNSE) Nanofabrication Facility at UC Riverside. The authors thank Dr. Krassimir Bozhilov (UCR) for his help with the HRTEM characterization and Dr. Pingrong Wei (UGA) for assistance with the single crystal x-ray diffraction measurements. S.R. acknowledges useful discussions with Dr. Valentine Kachorovskii (Ioffe Institute) and Dr. Michael Levinshtein (Ioffe Institute) on low-frequency noise in low-dimensional materials.

**Contributions:** A.A.B. conceived the idea, coordinated the project, led the experimental data analysis and manuscript preparation; A.K.G. fabricated the devices, conducted electrical and noise measurements, and contributed to data analysis; S.R. contributed to the noise data analysis; T.T.S. supervised material synthesis and contributed to materials characterization; M.A.B. synthesized the material and performed materials characterization. All authors contributed to the manuscript preparation.

## References

1    B. Sipos, A. F. Kusmartseva, A. Akrap, H. Berger, L. Forró and E. Tutiš, *Nature Materials*, 2008, **7**, 960.

2    K. S. Novoselov, A. Mishchenko, A. Carvalho and A. H. Castro Neto, *Science*, 2016, **353**, 461.

3    D. Jariwala, V. K. Sangwan, L. J. Lauhon, T. J. Marks and M. C. Hersam, *ACS Nano*, 2014, **8**, 1102–1120.

4    K. S. Novoselov, A. K. Geim, S. V Morozov, D. Jiang, Y. Zhang, S. V Dubonos, I. V Grigorieva and A. A. Firsov, *Science*, 2004, **306**, 666–669.

5    L. Li, Y. Yu, G. J. Ye, Q. Ge, X. Ou, H. Wu, D. Feng, X. H. Chen and Y. Zhang, *Nature Nanotechnology*, 2014, **9**, 372.

6    S. Z. Butler, S. M. Hollen, L. Cao, Y. Cui, J. A. Gupta, H. R. Gutiérrez, T. F. Heinz, S. S. Hong, J. Huang, A. F. Ismach, E. Johnston-Halperin, M. Kuno, V. V Plashnitsa, R. D. Robinson, R. S. Ruoff, S. Salahuddin, J. Shan, L. Shi, M. G. Spencer, M. Terrones, W. Windl and J. E. Goldberger, *ACS Nano*, 2013, **7**, 2898–2926.

7    A. K. Geim and I. V Grigorieva, *Nature*, 2013, **499**, 419.

8    A. A. Balandin, *Nature Materials*, 2011, **10**, 569.

9    X. Cui, G.-H. Lee, Y. D. Kim, G. Arefe, P. Y. Huang, C.-H. Lee, D. A. Chenet, X. Zhang, L. Wang, F. Ye, F. Pizzocchero and B. S. Jessen, *Nature Nanotechnology*, 2015, **10**, 534–540.

10   Y. Zhang, Y.-W. Tan, H. L. Stormer and P. Kim, *Nature*, 2005, **438**, 201.

11   B. Radisavljevic, A. Radenovic, J. Brivio, V. Giacometti and A. Kis, *Nature Nanotechnology*, 2011, **6**, 147.

12   P. Monceau, *Advances in Physics*, 2012, **61**, 325–581.

13   J. O. I. and A. J. M.-M. and M. B. and R. B. and E. F. and J. M. C. and J. R. A. and C. S. and H. S. J. van der Z. and R. D. and I. J. F. and A. Castellanos-Gomez, *2D Materials*,

2017, **4**, 22003.

14    J. Dai, M. Li and X. C. Zeng, *Wiley Interdisciplinary Reviews: Computational Molecular Science*, 2016, **6**, 211–222.

15    M. A. Stolyarov, G. Liu, M. A. Bloodgood, E. Aytan, C. Jiang, R. Samnakay, T. T. Salguero, D. L. Nika, S. L. Rumyantsev, M. S. Shur, K. N. Bozhilov and A. A. Balandin, *Nanoscale*, 2016, **8**, 15774–15782.

16    A. Geremew, M. A. Bloodgood, E. Aytan, B. W. K. Woo, S. R. Corber, G. Liu, K. Bozhilov, T. T. Salguero, S. Rumyantsev, M. P. Rao and A. A. Balandin, *IEEE Electron Device Letters*, 2018, **39**, 735–738.

17    J. Lienig, *Proceedings of the 2013 ACM International Symposium on Physical Design*, ACM, 2013, 33–40.

18    J. Gambino, T. C. Lee, F. Chen and T. D. Sullivan, *Proceedings of the International Symposium on the Physical and Failure Analysis of Integrated Circuits, IPFA*, 2009, 677–684.

19    A. A. Balandin, *Nature Nanotechnology*, 2013, **8**, 549.

20    R. H. Koch, J. R. Lloyd and J. Cronin, *Physical Review Letters*, 1985, **55**, 2487–2490.

21    M. V. Haartman, M. Ostling, *Low-Frequency Noise in Advanced MOS Devices*, Springer, Dordrecht, 2010.

22    W. Yang and Z. Çelik-Butler, *Solid-State Electronics*, 1991, **34**, 911–916.

23    V. P. Kunets, R. Pomraenke, J. Dobbert, H. Kissel, U. Müller, H. Kostial, E. Wiebicke, G. G. Tarasov, Y. I. Mazur and W. T. Masselink, *IEEE Sensors Journal*, 2005, **5**, 883–887.

24    J. Huh, D. C. Kim, A. M. Munshi, D. L. Dheeraj, D. Jang, G. T. Kim, B. O. Fimland and H. Weman, *Nanotechnology*, 2016, **27**, 385703.

25    P. H. Handel, *IEEE Electron Device Letters*, 1994, **41**, 2023.

26    P. Dutta and P. M. Horn, *Reviews of Modern Physics*, 1981, **53**, 497–516.

27    T.-M. Chen and A. M. Yassine, *IEEE Transactions on Electron Devices*, 1994, **41**, 2165–

2172.

28    L. K. J. Vandamme, *IEEE Transactions on Electron Devices*, 1994, **41**, 2176–2187.

29    B. Neri, A. Diligenti and P. E. Bagnoli, *IEEE Transactions on Electron Devices*, 1987, **34**, 2317–2322.

30    S. Beyne, K. Croes, I. De Wolf and Z. Tőkei, *Journal of Applied Physics*, 2016, **119**, 184302.

31    H. Furuseth, S. and Fjellvag, Acta Chem. Scand.,1991, **45**, 694–697.

32    E. Canadell, Y. Mathey and M. H. Whangbo, *Journal of the American Chemical Society*, 1988, **110**, 104–108.

33    K. Sto and F. R. Wagner, *Journal of Solid State Chemistry*, 1998, **168**, 160–168.

34    R. Seshadri, E. Suard, C. Felser, E. W. Finckh, A. Maignan and W. Tremel, *Journal of Materials Chemistry*, 1998, **8**, 2869–2874.

35    C. Felser, E. W. Finckh, H. Kleinke, F. Rocker and W. Tremel, *Journal of Materials Chemistry*, 1998, **8**, 1787–1798.

36    K. Yamaya, S. Takayanagi and S. Tanda, *Physical Review B*, 2012, **85**, 184523.

37    M. Hoesch, G. Garbarino, C. Battaglia, P. Aebi and H. Berger, *Physical Review B*, 2016, **93**, 125102.

38    S. T. and K. M. and K. Y. and S. T. and S. T. and Y. Uwatoko, *New Journal of Physics*, 2017, **19**, 63004.

39    A. M. Ganose, L. Gannon, F. Fabrizi, H. Nowell, S. A. Barnett, H. Lei, X. Zhu, C. Petrovic, D. O. Scanlon and M. Hoesch, *Physical Review B*, 2018, **97**, 155103.

40    X. Zhu, W. Ning, L. Li, L. Ling, R. Zhang, J. Zhang, K. Wang, Y. Liu, L. Pi, Y. Ma, H. Du, M. Tian, Y. Sun, C. Petrovic and Y. Zhang, *Scientific Reports*, 2016, **6**, 26974.

41    S. Takahashi, T. Sambongi, J. W. Brill and W. Roark, *Solid State Communications*, 1984, **49**, 1031–1034.

42    M. Abdulsalam and D. P. Joubert, *The European Physical Journal B*, 2015, **88**, 177.

43    S. V Zaitsev-Zotov, *Microelectronic Engineering*, 2003, **69**, 549–554.

44    E. Slot, M. A. Holst, H. van der Zant and S. Zaitsev-Zotov, *Physical review letters*, 2004, **93**, 176602.

45    G. Liu, S. Rumyantsev, M. S. Shur and A. A. Balandin, *Applied Physics Letters*, 2013, **102**, 93111.

46    M. Zahid Hossain, S. Rumyantsev, M. S. Shur and A. A. Balandin, *Applied Physics Letters*, 2013, **102**, 153512.

47    J. Renteria, R. Samnakay, S. L. Rumyantsev, C. Jiang, P. Goli, M. S. Shur and A. A. Balandin, *Applied Physics Letters*, 2014, **104**, 153104.

48    S. L. Rumyantsev, C. Jiang, R. Samnakay, M. S. Shur and A. A. Balandin, *IEEE Electron Device Letters*, 2015, **36**, 517–519.

49    G. Liu, S. Rumyantsev, M. A. Bloodgood, T. T. Salguero and A. A. Balandin, *Nano Letters*, 2018, **18**, 3630–3636.

50    G. Liu, S. Rumyantsev, M. A. Bloodgood, T. T. Salguero, M. Shur and A. A. Balandin, *Nano Letters*, 2017, **17**, 377–383.

51    V. Mitin, L. Reggiani and L. Varani, in *Noise and Fluctuations Control in Electronic Devices, ed A. A. Balandin*, World Scientific Publishing Co., 2nd ed, 2003, Vol. 3, pp. 1–19.

52    A. A. Balandin, *Noise and Fluctuations Control in Electronic Devices*, World Scientific Publishing Co., 2003.

53    M. E. Levinshtein and S. L. Rumyantsev, *Semiconductor Science and Technology*, 1994, **9**, 1183.

54    R. B. Hall, *Thin Solid Films*, 1971, **8**, 263–271.

55    L. D. Landau and E. M. Lifshitz, in *Electrodynamics of Continuous Media*, Elsevier, Amsterdam, 2nd ed, 1984, vol. 8, ch. 1, pp. 1–33.

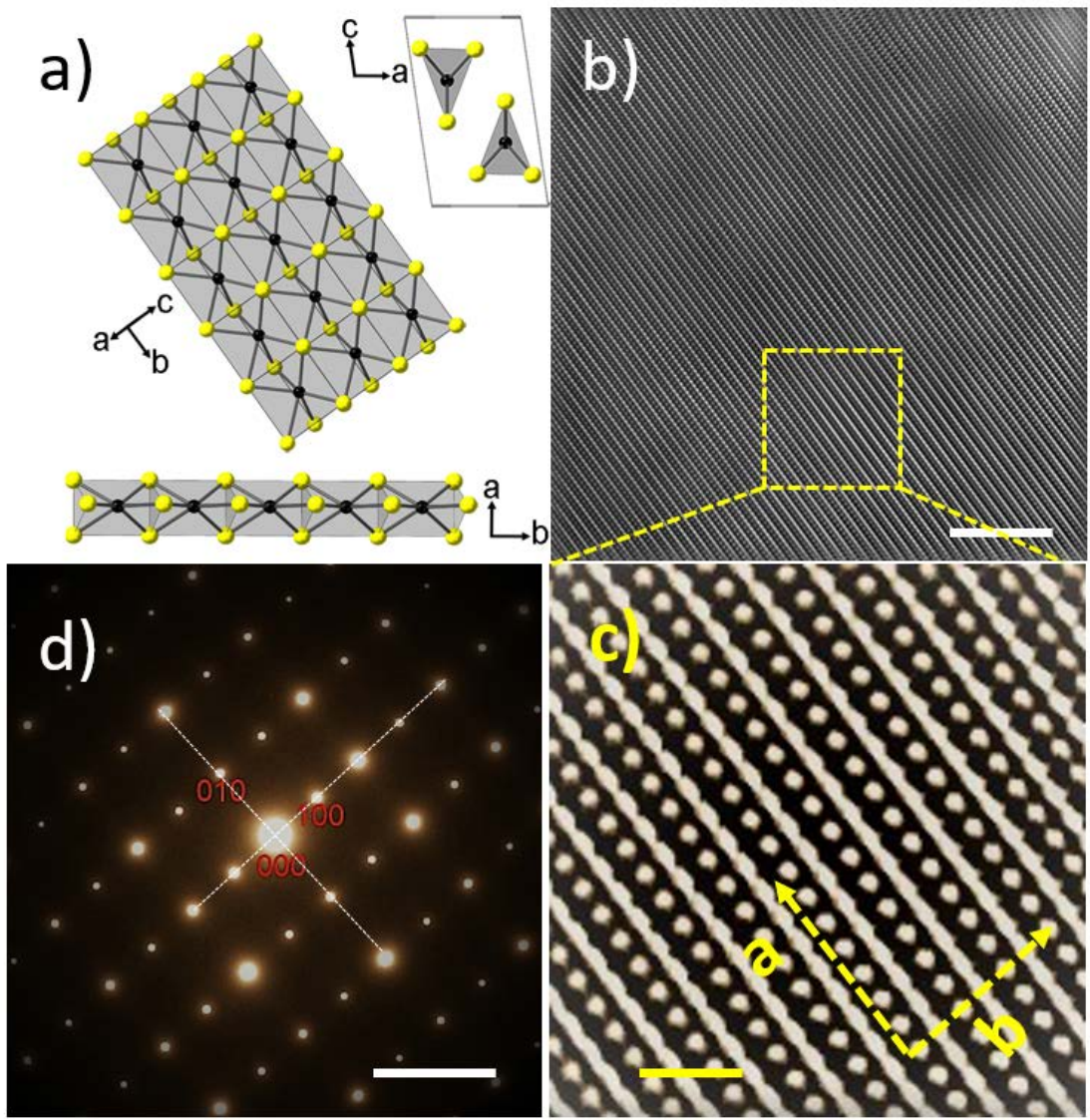

**Figure 1**: (a) Schematic of the structure of $ZrTe_3$. (b) HRTEM of a $ZrTe_3$ nanoribbon that shows a representative defect-free region. (c) HRTEM of $ZrTe_3$ with the orientation of atoms illustrated by the array of columns in panel a. (d) Selected area electron diffraction pattern along the (001) zone axis. This pattern was used to determine the lattice constants in the unit cell of $ZrTe_3$, which were consistent with the single crystal x-ray structure. The scale bar for (b), (c) and (d) are 7 nm, 1 nm and, 1 nm$^{-1}$ respectively.

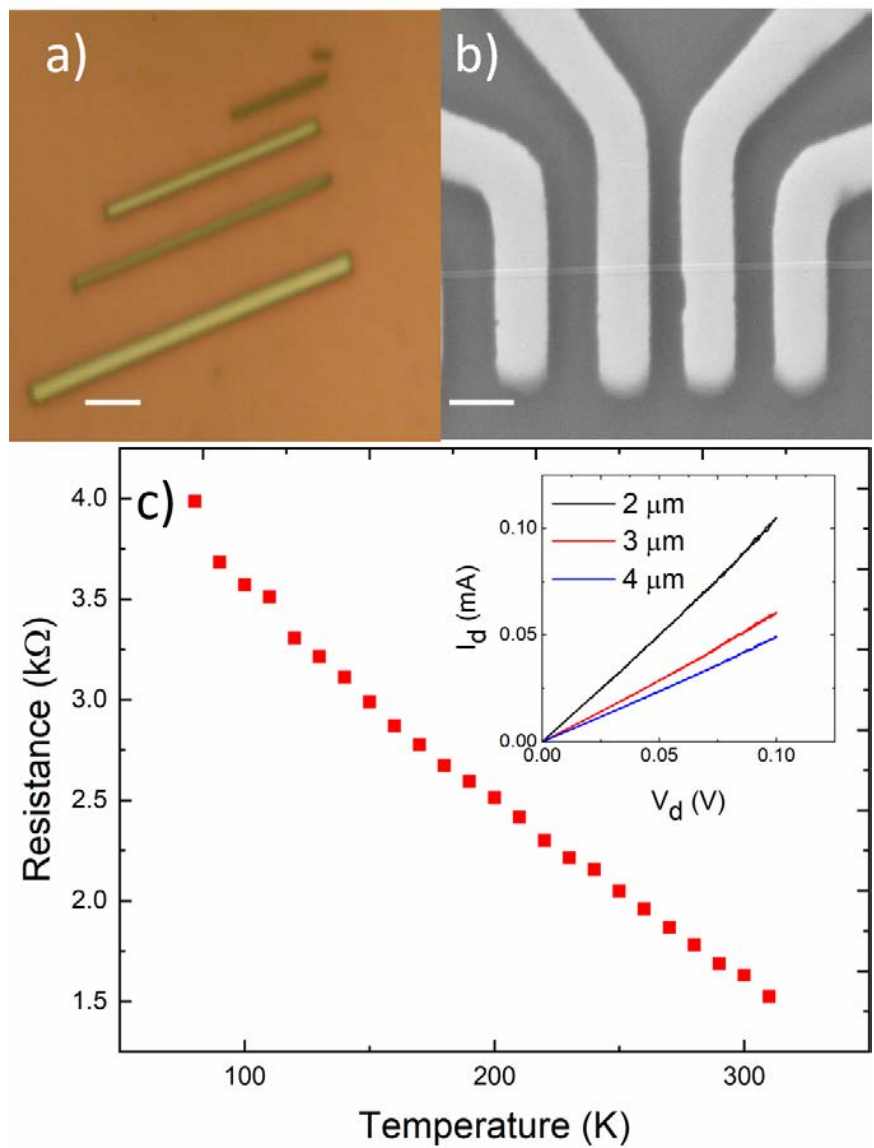

**Figure 2**: (a) Optical microscopy image of the mechanically exfoliated ZrTe$_3$ ribbons on SiO$_2$/Si substrate, before the contact evaporation. Different colors, green *vs.* gold, indicate the variation in the thickness of the nanoribbons. The scale bar is 1 μm. (b) Scanning electron microscopy image of a device fabricated from the quasi-1D ZrTe$_3$ nanoribbon with the thickness of ~50 nm. The scale bar is 2 μm. (c) The resistance of the quasi-1D ZrTe$_3$ nanoribbon as a function of temperature. The resistance shows semiconducting behavior at low temperature. The inset shows low-field I-Vs for representative devices of different lengths, as proof that the fabricated contacts were Ohmic.

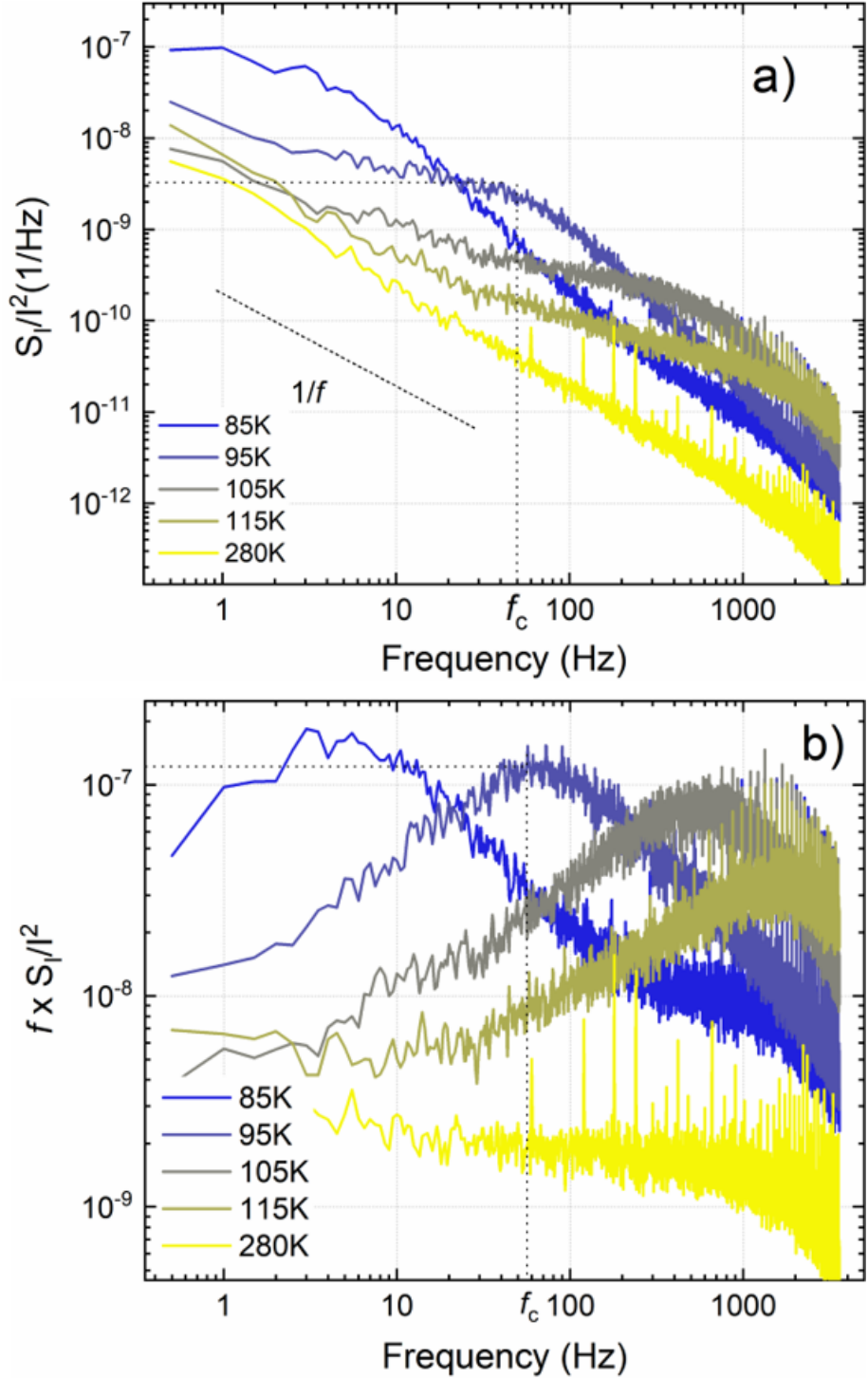

**Figure 3**: (a) Normalized noise spectral density, $S_I/I^2$, as a function of frequency of quasi-1D ZrTe$_3$ nanoribbon at temperatures from 85 K to 280 K. (b) Normalized noise spectral density multiplied by frequency, $S_I/I^2 \times f$, as a function of frequency. The bias voltage is V$_D$=0.1 V. The position of the characteristics frequency, $f_c$, is shown in broken line.

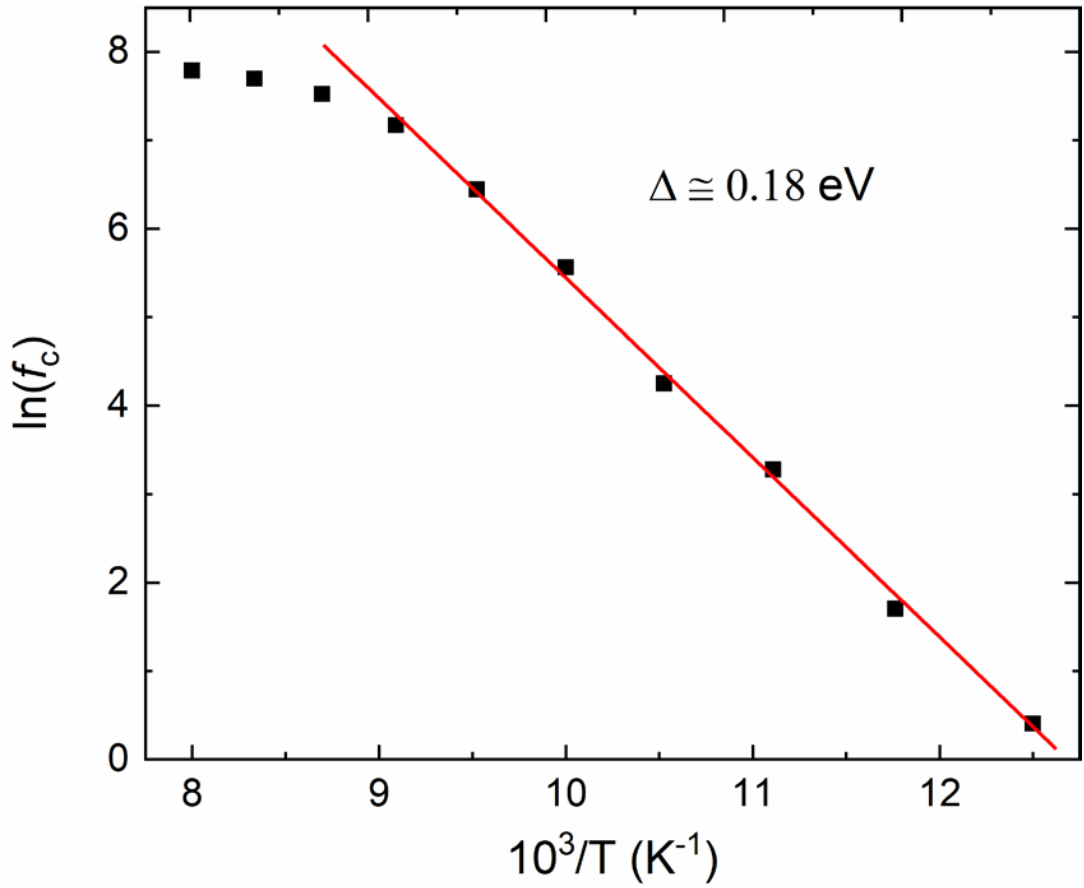

**Figure 4:** Arrhenius plot of the characteristics frequency, ln($f_c$), as a function of the inverse temperature, in quasi-1D ZrTe$_3$ device.

Adane K. Geremew, Sergey Rumyantsev, Matthew A. Bloodgood, Tina T. Salguero and Alexander A. Balandin - 2018

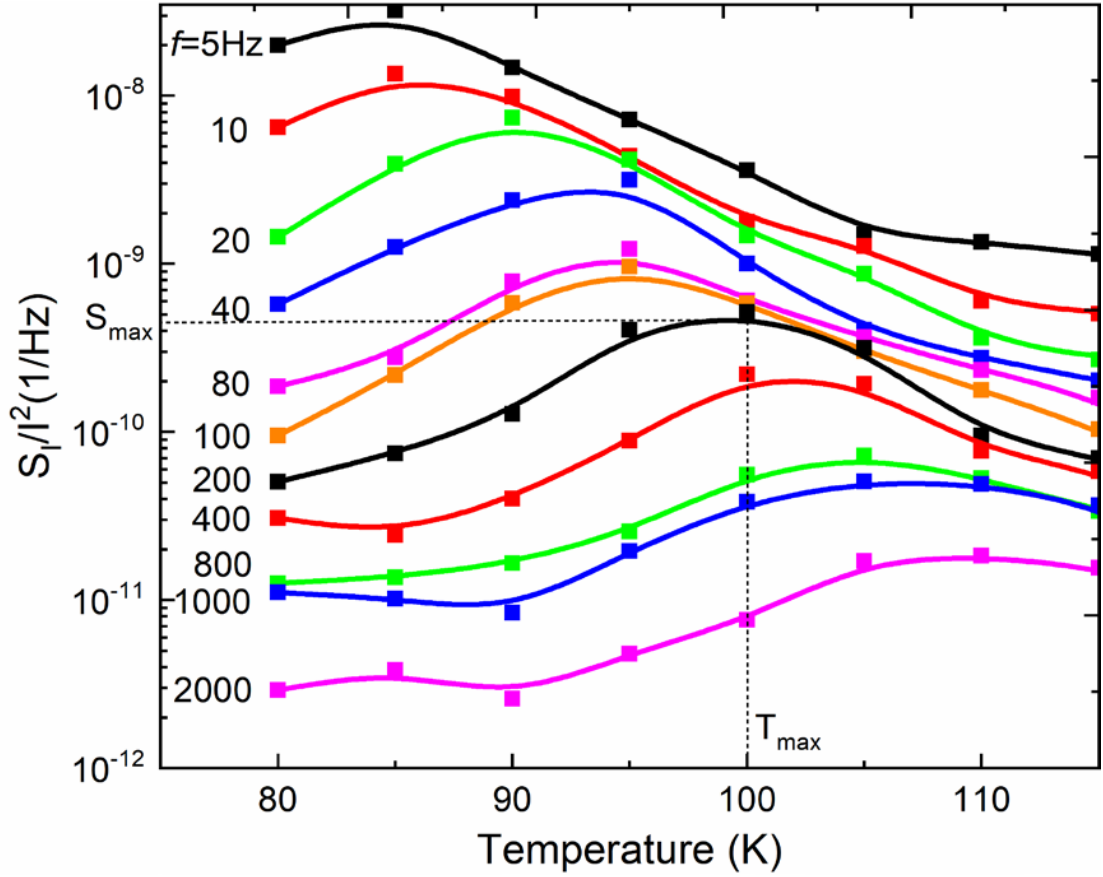

**Figure 5**: Normalized noise spectral density, $S_I/I^2$, as a function of temperature for different frequencies. As the temperature increases, the noise peak shifts to higher temperatures.

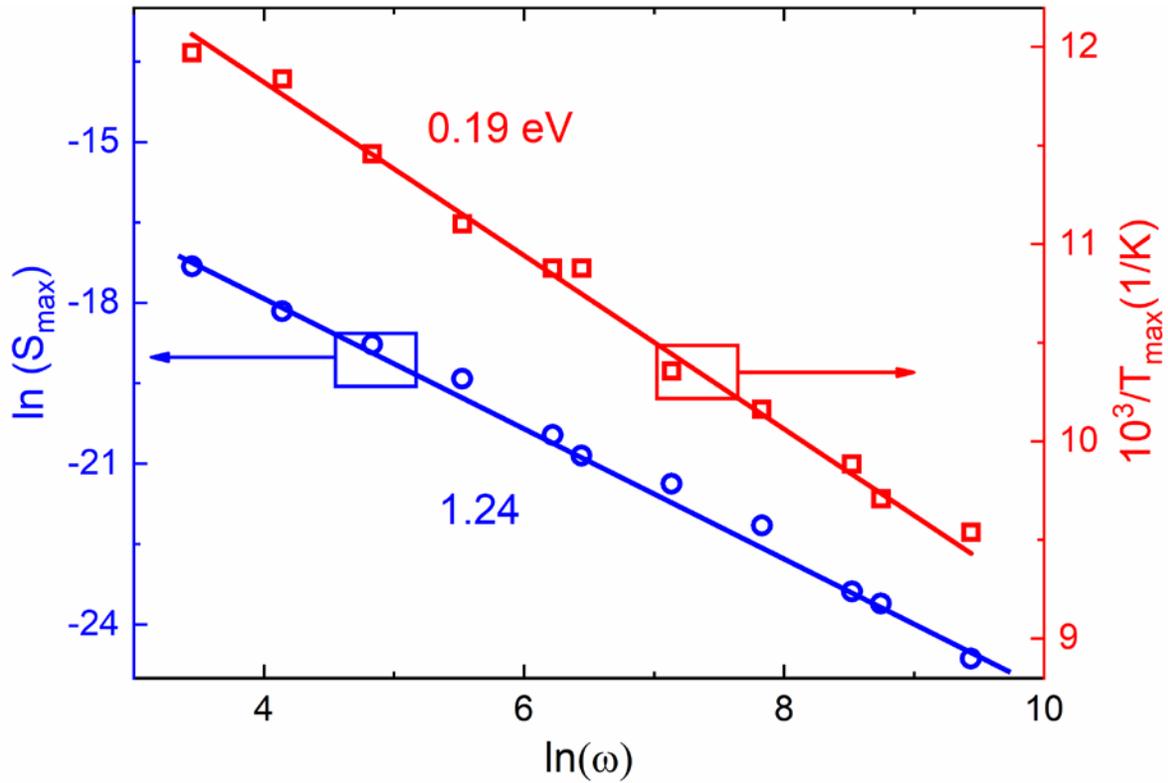

**Figure 6:** Maximum noise peak position, $S_{max}$, as a function of frequency (rad/sec) (blue curve). The extracted slope of the linear fitting is 1.24. The Arrhenius plot of the invers temperature ($K^{-1}$) as a function of frequency (rad/sec) (red curve). The extracted activation energy from the linear fitting is 0.19 eV.

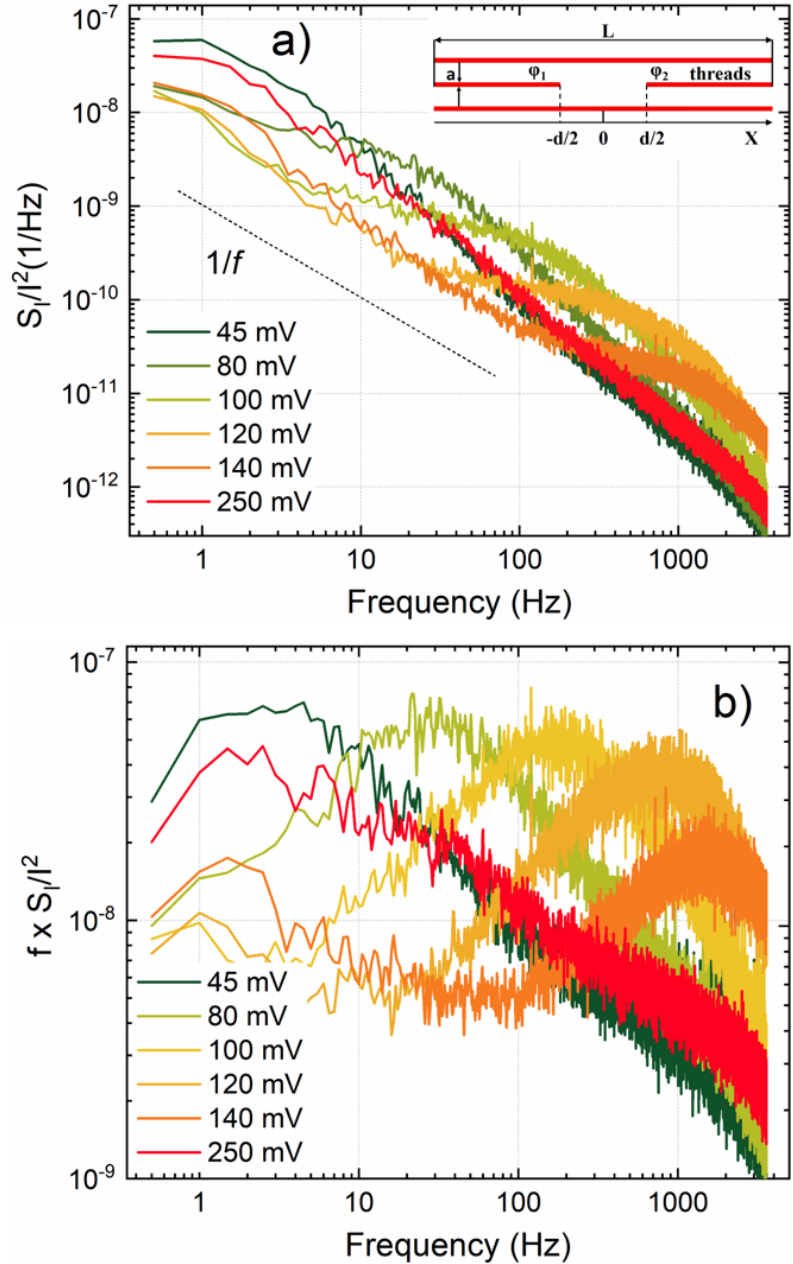

**Figure 7**: (a) Normalized noise spectral density, $S_I/I^2$, as a function of frequency of quasi-1D ZrTe$_3$ nanoribbon at the bias voltage ranging from 45 mV to 250 mV. Inset shows a schematic of the atomic thread bundle with one broken thread, illustrating the action of a defect. (b) Normalized noise spectral density multiplied by frequency, $S_I/I^2 \times f$, as a function of frequency. Note an unusually strong dependence of the Lorentzian peak on the bias voltage. The data are presented for T = 78 K.